\documentclass[twocolumn,superscriptaddress,showpacs,showkeys,prl]{revtex4-1}
\usepackage{amsmath}
\usepackage{graphicx}

\newlength {\defaultparindent }
\newenvironment {annotation Text}{}{}
\newcommand\includegraphicss[1]
  {\resizebox{6.5cm}{!}{\includegraphics[clip=true]{#1}}}
\begin{document}

%\title{Switching reciprocity on and off in a nuclear resonance forward
%scattering experiment}
\title{Large reciprocity violation switched on and off in a nuclear
resonance
% forward
scattering experiment}
\author{L. De\'{a}k}
\email{deak.laszlo@wigner.mta.hu}
\affiliation{Wigner\ RCP, RMKI, P.O.B. 49, H-1525 Budapest, Hungary}
\author{L. Botty\'{a}n}
\affiliation{Wigner\ RCP, RMKI, P.O.B. 49, H-1525 Budapest, Hungary}
\author{T. F\"{u}l\"{o}p}
\affiliation{Wigner\ RCP, RMKI, P.O.B. 49, H-1525 Budapest, Hungary}
\author{G. Kert\'{e}sz}
\affiliation{Wigner\ RCP, RMKI, P.O.B. 49, H-1525 Budapest, Hungary}
\author{D. L. Nagy}
\affiliation{Wigner\ RCP, RMKI, P.O.B. 49, H-1525 Budapest, Hungary}
\author{R. R\"{u}ffer}
\affiliation{European Synchrotron Radiation Facility, BP 220, F--38043 Grenoble, France}
\author{H. Spiering}
\affiliation{%
%Institut f\"{u}r Anorganische und Analytische Chemie, 
Johannes Gutenberg
Universit\"{a}t Mainz, Staudinger Weg 9, D-55099 Mainz, Germany}
\author{F. Tanczik\'{o}}
\affiliation{Wigner\ RCP, RMKI, P.O.B. 49, H-1525 Budapest, Hungary}
\author{G. Vank\'{o}}
\affiliation{Wigner\ RCP, RMKI, P.O.B. 49, H-1525 Budapest, Hungary}
\date{\today }

 \begin{abstract}
Reciprocity is when the scattering amplitude of wave propagation
satisfies a symmetry property, connecting a scattering process with an
appropriate reversed one. We report on an experiment using nuclear 
resonance scattering of synchrotron radiation, which demonstrates 
that magneto-optical materials do not necessarily violate reciprocity. 
The setting enables to switch easily between reciprocity and its violation. 
In the latter case, the exhibited reciprocity violation is orders 
of magnitude larger than achieved by previous wave scattering experiments.
%enabled to switch reciprocity on and off and, in the latter case, to
%produce large reciprocity violation, orders of magnitude larger than
%achieved by previous experiments.
 \end{abstract}

\keywords{reciprocity, scattering theory, nuclear resonance scattering,
M\"{o}ssbauer spectroscopy}
\pacs{03.65.Nk, 76.80.+y, 78.70.Ck, 78.20.Ls}
\maketitle
\preprint{HEP/123-qed}  %% !!!

The reciprocity principle, which states that the interchange of source
and detector does not change the scattering amplitude of a wave
scattering process, cannot be derived from first principles and is not
necessarily fulfilled. The term `reciprocity' has been introduced by
\textcite{Stokes1849}, and the numerous subsequent related publications
cover the whole 20$^{\mathrm{th}}$ century, as it is summarized in the
review paper of \textcite{Potton2004}. Reciprocity theorems were derived
for various scattering problems, telling under which conditions and
limitations the reciprocity principle is valid
\cite{deHoop,Saxon1955,Bilhorn1964,Carminati2000,Hillion1978}, and
situations in the field of local and nonlocal electromagnetism 
\cite{Helmholtz1866,Lorentz1905,Xie2009,Xie2009b}, sound waves
\cite{Strutt1877}, electric circuits \cite{Carson1924}, radio
communication \cite{Carson1930}, and local and nonlocal quantum
mechanical scattering problems \cite{Bilhorn1964,Xie2008,Leung2010} were
considered. Nonreciprocal devices (circulators and isolators) with
on-chip integration possibility were also suggested \cite{Kamal2011}. In
a recent work of \textcite{DeaFul11}, a general reciprocity theorem was
formulated, which covers all cases of wave phenomena that can be
represented by a Schr\"odinger equation, with Hamiltonian $H=H_{0}+V$,
where $H_{0}$ describes free wave propagation and $V$ the scatterer.
Reciprocity is more general than time reversal invariance, can occur for
absorptive scattering media as well, and it also fundamentally differs
from rotational invariance \cite{DeaFul11}. We note that, in X-ray
optics, the term non-reciprocity may also refer to time-reversal odd
optical activity 
\cite{vanderLaan2001,Goulon1999,Goulon2000,Goulon2003,Matsubara2005},
which meaning differs from the historical one used here
\cite{Potton2004,DeaFul11}.

Here, we report on an experimental investigation of reciprocity and its
violation. Our aim was threefold: to show that magneto-optical materials
do not necessarily violate reciprocity, to control easily whether
reciprocity is present or missing, and to demonstrate that reciprocity
violation can be a remarkably strong effect. The example considered
belongs to the field of optics, where the multiple scattering of X-rays
or neutrons can be described by an index of refraction \cite{Lax51}. In
forward scattering geometry, \textcite{Blume68} established, already in
1968, the theory for M\"{o}ssbauer absorption of $\gamma$ radiation. The
theory was later extended for  grazing incidence scattering on
stratified media  \cite{Hannon85b,Deak96,Roehlsberger2003,Andreeva2005}
and computer programs also became available
\cite{Sturhahn1994,Shvydko2000,Spiering00,Andreeva2008,Sajti2009}. As a
result, reciprocity situations can be simulated and the condition and 
aspects of reciprocity can be tested.
  \noindent
%\begin{figure}[p]
\begin{figure}[t]
%\resizebox{.45\columnwidth}{!}{\includegraphics{Fig1abc.eps}}
\includegraphicss {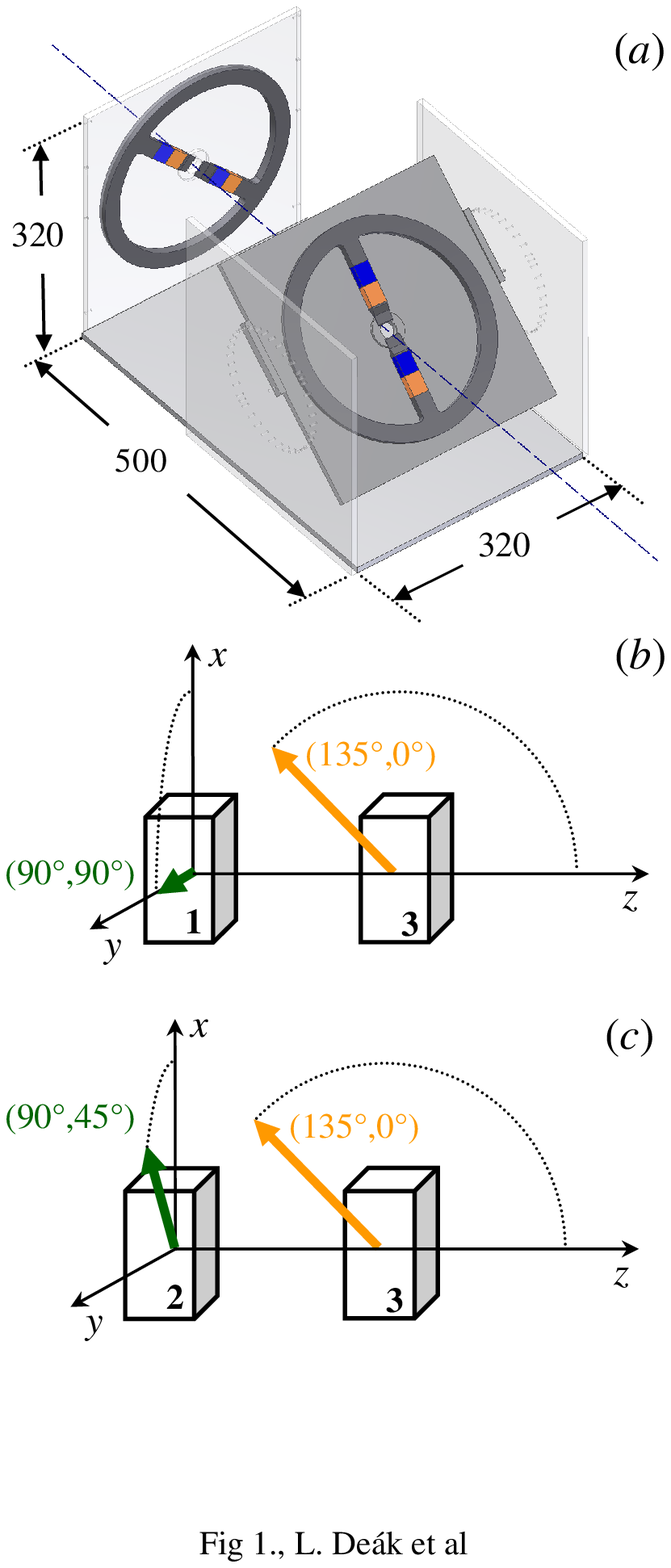}
 \caption{(color online). (\textit{a}) Sample holder for the reciprocity
test of nuclear resonance scattering of synchrotron radiation (sizes in
units of mm). The dashed line indicates the path of the synchrotron beam
going through two slits of 0.5\ mm diameter. The two $\protect \alpha
-^{57}$Fe foils are mounted on the slits between permanent magnets
providing 0.19\ T magnetic field in the required direction, adjustable
by the dark grey wheels mounted in the plane of the foils. (\textit{b}),
(\textit{c}) Geometrical arrangement for the (magnitude) reciprocal and
the nonreciprocal case, respectively. The synchrotron beam propagates in
the direction of the $z$ axis and gets scattered on foils 1, 3
(reciprocal scattering) or 2, 3 (nonreciprocal scattering). The thick
arrows show the direction of the magnetic field in the iron foils, also
given by the polar angles.
 }
 \label{Fig1abc}
\end{figure}

The experimental method is nuclear resonance scattering of
synchrotron radiation that was
introduced by \textcite{Gerdau1985}, an analog of classical M\"{o}ssbauer
spectroscopy. The experiment was performed in forward-scattering geometry 
\cite{Hastings1991} corresponding to a conventional 
transmission M\"{o}ssbauer arrangement. In contrast to laboratory 
M\"{o}ssbauer experiments performed in energy domain, nuclear resonance 
scattering experiments are performed in the time domain, i.e. the time 
response of the scatterer to the excitation by a synchrotron pulse is 
recorded and analyzed. The synchrotron radiation bunch excites the 
hyperfine-split nuclear energy levels simultaneously, leading to 
characteristic beats in the time spectra. The scatterer consisted of 
two foils mounted in a sample holder, each foil being a $6\ \mathrm{\mu}$m 
thick ferromagnetic $^{57}$Fe absorber uniformly magnetized in a field of 
$0.19\ \mathrm{T}$ of permanent magnets as shown in Fig.~1a.

We considered two experimental arrangements of the scatterer differing 
from each other in the directions of the magnetic fields acting on the 
iron foils according to Figs.~1b and 1c that we shall refer to, henceforth, 
as arrangements 1b and 1c, respectively. For both arrangements of the
scatterer, the scattering time response (hereafter called 'time spectrum') 
was compared to that of the reciprocal scattering. The interchange of 
source and detector position was realized, following \cite{DeaFul11}, 
by a $180^{\circ}$ rotation of the whole sample holder around the 
$x$ axis of the coordinate system indicated in the figure. 

The experiment was performed at the Nuclear Resonance
side-station ID22N of the European Synchrotron Radiation Facility (ESRF) 
delivering $\sigma$- (i.e., horizontally) polarized beam. The detector was 
a Si avalanche photo diode (APD) in front of which a Si(840) channel 
cut $\sigma$-analyzer was applied. Since in this particular case both the 
source and the detector were of $\sigma$-polarization, the reciprocal 
scattering could be realized by a mere $180^{\circ}$ rotation of the 
scatterer. The four time spectra (count vs.\ time diagrams) shown in 
Figs.~2 and 3 were measured with the respective arrangements 1b and 1c 
of the scatterer. Spectra (a) and (b) in each figure correspond to the 
direct and the reciprocal scattering geometry, respectively. Clearly, arrangement 
1b provided an example where the measured intensity was reciprocal 
while arrangement 1c showed a large violation of reciprocity. 
How was it possible to find such scatterer arrangements?

In scalar wave phenomena, the interchange of source and detector defines
the reciprocal process uniquely but, for waves with more than one
spin/polarization component, the polarizations of the reciprocal
process must be chosen appropriately to obtain reciprocity.
The conventionally used condition of reciprocity in linear systems
\cite{Potton2004} is the self-transpose (also called complex symmetric)
property of the matrix of the scattering potential, of the index of
refraction, of the dielectric/magnetic permeability tensors, or of the
Green's function \cite{Potton2004,Xie2008,Xie2009}. This matrix must be
considered in the polarization basis distinguished by the scattering
processes in question \cite{DeaFul11}.
For our scatterers, the potential $V$ is zero in vacuum and, within
each iron foil, is a space independent operator that depends on the
direction $(\theta,\phi)$ of the hyperfine magnetic field in the layer.
The incoming synchrotron beam is $\sigma$-polarized and the same
polarization is measured by the analyzer, and it is the $\sigma, \pi$
polarization basis in which self-transposeness of the scattering
potential matrix $V$ ensures reciprocity. For the $\Delta m = -1$
(M\"ossbauer) transition under discussion \cite{Note1}, this matrix
reads
 \begin{align}
\label{eaa}V = 2c \left(
\begin{array}
[c]{cc}%
1 - \sin^{2} \phi \sin^{2}\theta &
 -\frac{\sin^{2}\theta \sin2\phi}{4} + i\cos \theta \\
-\frac{\sin^{2}\theta \sin2\phi}{4} - i\cos \theta
& 1 - \cos^{2} \phi \sin^{2}\theta
\end{array}
\right)  \allowbreak
 \end{align}
% \begin{align}
%\label{eaa}V = 2c
%\hskip-.26ex
% \left(
%\begin{array}
%[c]{cc}%
%1 - \sin^{2} \phi \sin^{2}\theta &
% i\cos \theta - \frac{\sin^{2}\theta \sin2\phi}{4}
%\hskip-.3ex
%\\
%\hskip-.3ex
% - i\cos \theta - \frac{\sin^{2}\theta \sin2\phi}{4}
%& 1 - \cos^{2} \phi \sin^{2}\theta
%\end{array}
%\right)
%\!
% \end{align}
in a given layer with magnetic field direction $(\theta,\phi)$, where
the complex coefficient $c$ is the same in both layers \cite{DeaFul11}.

Let us consider three possible orientations for the in-layer
magnetization, and the corresponding potentials,
 \begin{align}
\label{eab}\left(  \theta_{1}, \phi_{1} \right)   &  = \left(  90^{\circ},
90^{\circ} \right) , & V_{1}  &  = c\left(
\begin{array}
[c]{cc}%
0 & 0\\
0 & 2
\end{array}
\right)  ,\\
\left(  \theta_{2}, \phi_{2} \right)   &  = \left(  90^{\circ}, 45^{\circ}
\right) , & V_{2}  &  = c\left(
\begin{array}
[c]{cc}%
1 & -\frac{1}{2}\\
-\frac{1}{2} & 1
\end{array}
\right)  ,\\
\left(  \theta_{3}, \phi_{3} \right)   &  = \left(  135^{\circ}, 0^{\circ}
\right) , & V_{3}  &  = c\left(
\begin{array}
[c]{cc}%
2 & -i\sqrt{2}\\
i\sqrt{2} & 1
\end{array}
\right)  .
 \end{align}
Apparently, $V_{1}$ and $V_{2}$ are self-transpose, and $V_{3}$ is not.

A scatterer consisting of two layers with one foil having $V_{1}$ and
the other having $V_{2}$ would mean a self-transpose potential and would
obviously provide reciprocity. Therefore, so as to study nontrivial
cases, in our experiment, arrangement 1b of the scatterer was a combination of a
$V_{1}$ layer and of a $V_{3}$ one, while arrangement 1c of the scatterer was
formed by a $V_{2}$ foil and a $V_{3}$ one. Neither combination is
self-transpose so both situations are nonreciprocal.

The obtained experimental spectra can be seen in Figs.~2 and 3,
respectively, together with the theoretical simulation predictions. We
can see that the combination of $V_{1}$ and $V_{3}$ (i.e., arrangement 1b) 
results in reciprocity in the measured intensities, while the pair of foils with
$V_{2}$ and $V_{3}$ (i.e., arrangement 1c) exhibits apparent nonreciprocity. 
The latter outcome is remarkable because of the large nonreciprocal effect: 
The ratio of intensities of direct and reciprocal scattering is
% almost
nearly
$10^{3}$ in certain time intervals.
\noindent
%\begin{figure}[p]
\begin{figure}[t]
 \includegraphicss {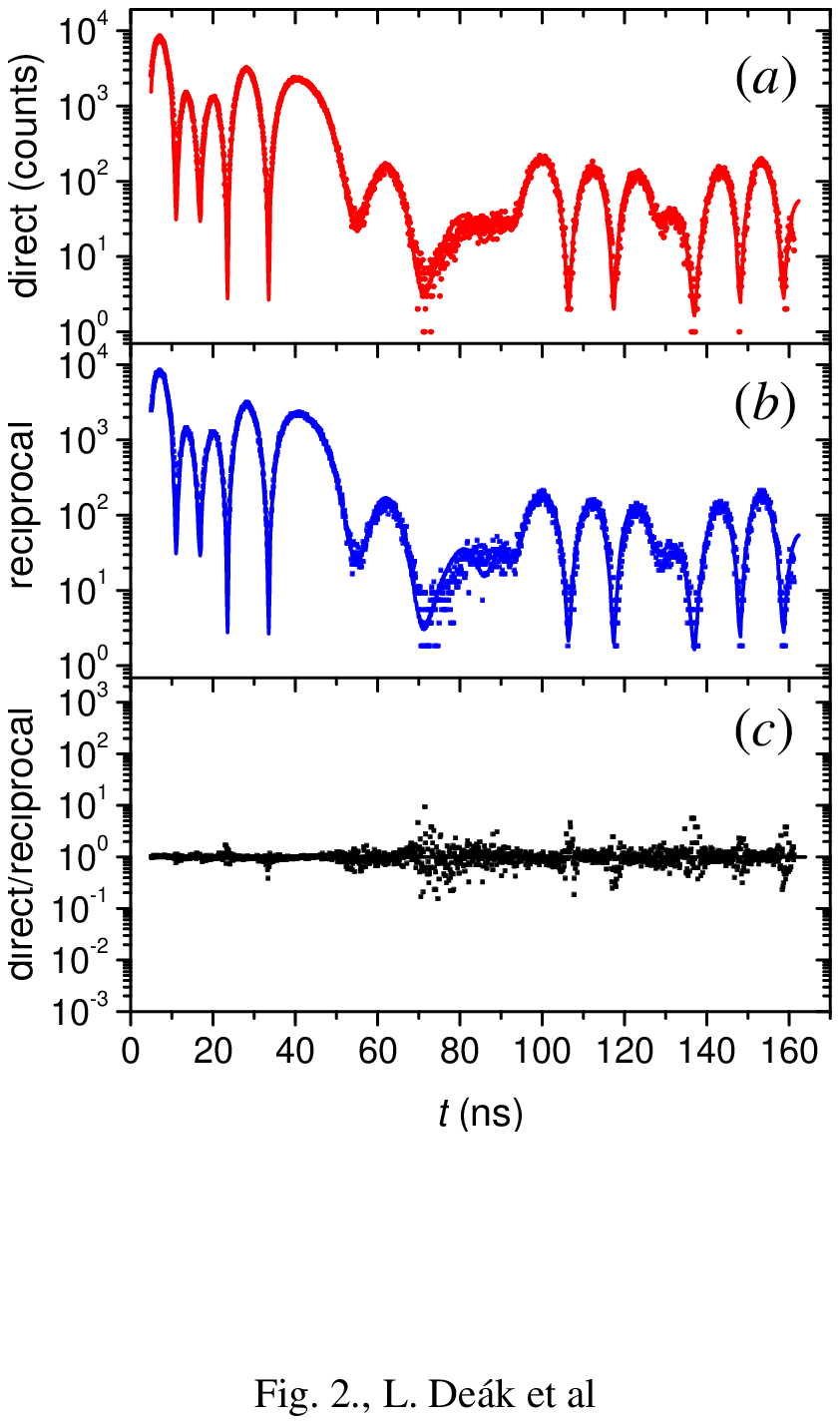}
%\makebox[\columnwidth]
%  {\resizebox{.47\columnwidth}{!}{\includegraphics{Fig2abc.eps}}
%\hfill
%  \resizebox{.47\columnwidth}{!}{\includegraphics{Fig3abc.eps}}}

\caption{(color online). (\textit{a}) Measured (dotted lines) and
simulated (solid lines) nuclear resonance forward scattering of
synchrotron radiation time spectra  on two $\mathrm{\protect \alpha
}-^{57}\mathrm{Fe}$ foils of $6\  \mathrm{\protect \mu m}$ thickness
using the polarizer-analyser setup for incident $\protect \sigma $
(\textit{viz}. horizontal electric) linearly polarized photons
scattered to the same polarization ($\protect \sigma \rightarrow
\protect \sigma $ scattering) for the case of hyperfine magnetic field
orientations shown in Fig.~1b (foils 1, 3). (\textit{b}) Results for the
reciprocal (source-detector exchanged) situation realized by the
180${}^\circ$ rotation of the scatterer (Fig.~1a). (\textit{c}) The
amount of nonreciprocity, displayed as the ratio of the counts for the
direct process \textit{(a)} and of the reciprocal process \textit{(b)}.
The ratio is found to be near the constant 1, which is the value when
the intensities exhibit reciprocity. Slight deviations of the ratio from
1 can be seen in the regions of the minima of the beatings, where the
count rates are low and statistical errors are significant.
 }
 \label{Fig2abc}
\end{figure}

%\begin{figure}[p]
\begin{figure}[t]
 \includegraphicss {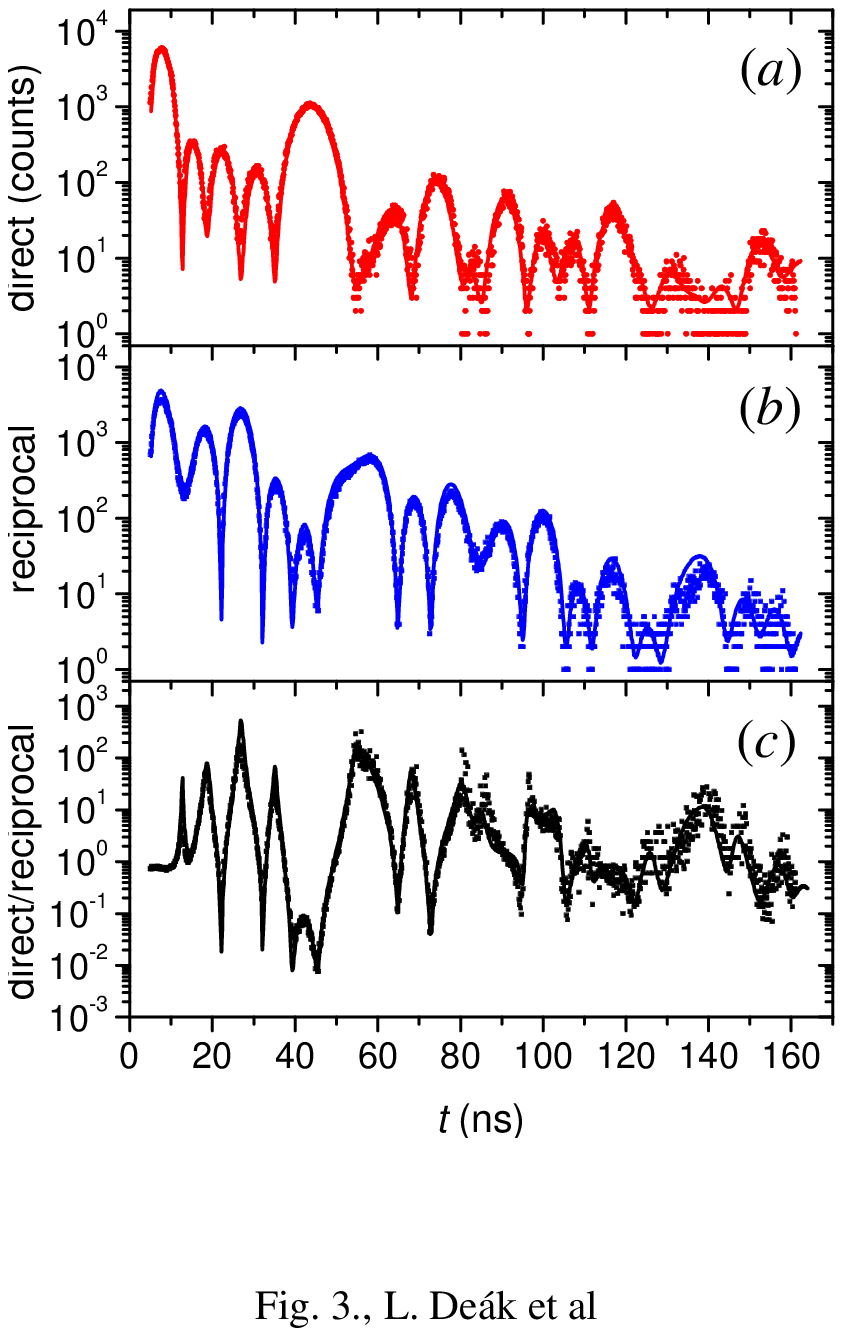}
%  \resizebox{.47\columnwidth}{!}{\includegraphics{Fig3abc.eps}}
\caption{(color online). Results for the hyperfine magnetic field orientations 
shown in Fig.~1c (foils 2, 3)---figure analogous to Fig.~2. Up to three orders
of magnitude large nonreciprocity ratio is observed.}
 \label{Fig3abc}
\end{figure}

In parallel, the former result seems surprising since we have just found
that reciprocity must be violated for that setting. The explanation is
that, in experiments that detect probabilities/counts, i.e., only the
squared absolute value of the complex scattering amplitude (i.e., intensity), 
reciprocity violation remains hidden if it is present only in the phase. Ref.\
\onlinecite{DeaFul11} calls such cases magnitude reciprocity, and proves
that, if there is a space independent angle $\delta$ such that 
 \begin{equation}
\label{ead}V_{12}(\mathbf{r}) = e^{i \delta} V_{21}(\mathbf{r})
 \end{equation}
at any position $\mathbf{r}$,
then magnitude reciprocity takes place. Applying this condition to our
two scatterers, the first one (layers $V_{1}$ and $V_{3}$) proves to be
a case of magnitude reciprocity---the phase factor ``repairing'' $V_{3}$
obviously ``repairs'' $V_{1}$ as well. Actually, our intention with this
scatterer was to provide an example for the phenomenon of magnitude
reciprocity, to drive attention to this weaker but important version of
reciprocity. On the other side, ``repairing'' $V_{2}$ and $V_{3}$
requires two different phase factors so there is no common phase factor
for them.

The experimental results and the corresponding computer simulations made
by the computer program \cite{Spiering00} are both shown in Figs.~2, 3.
The slight imperfection in the agreement between measurement and
simulation is due to, and informs about, nonperfect uniformness in the
foil thickness and the magnetic field inside. The experiment kept these
undesired influences under control by using the high collimation and
brilliance of the synchrotron beam, which allowed the usage of slits of as
small as $0.5\  \mathrm{mm}$ of width selecting adequate homogeneous
parts of both foils being $39.2\  \mathrm{cm}$ far from each others. The
small size of the slits ensures that, after the $180^{\circ}$ rotation
of the sample holder, the same part of the foils is illuminated. The
agreement of the experimental spectrum with that of the magnitude
reciprocal counterpart setting---seen in Fig.~2---justifies the
confidence that the recipocal situation was achieved to a high accuracy.

In summary, we have realized a (magnitude) reciprocal and a
nonreciprocal experimental arrangement of magnetized $\alpha$-
${}^{57}\mathrm{Fe}$ foils, which had neither time reversal invariance
nor $180^{\circ}$-rotational symmetry. Using nuclear resonance scattering
of synchrotron radiation, depending on the easily adjustable
experimental geometry, reciprocity, and also three orders of magnitude
large nonreciprocity, was experimentally observed in the intensities, in
full agreement with the theoretical expectations. The presence of
magneto-optic Faraday effect does not automatically lead to
nonreciprocity. Further applications in the field of $\gamma$-optics are
expected, as nonreciprocal devices belong to an important class of
optical components.

 \begin{acknowledgments}
This work was partly supported by the Hungarian Scientific Research Fund
(OTKA), the National Office for Research and Technology of Hungary
(NKTH), and the European Research Council under contract numbers K81161,
NAP-Veneus'08, and StG-259709, respectively. The authors gratefully
acknowledge the beam time supplied by the European Synchrotron Radiation
Facility (ESRF) for experiment SI-1794. The authors are thankful to Ilya
Sergeev (ESRF) and Csilla Bogd\'an (Wigner RCP) for technical
assistance.
 \end{acknowledgments}

%\bibliographystyle{apsrev4-1}
%\bibliography{acompat,exprecibib-120423}

\newif\ifabfull\abfulltrue

\end{document}